\documentclass[aps,floatfix,notitlepage,groupedaddress,superscriptaddress,amsmath,amssymb,singlecolumn,jpa]{revtex4-1}

\usepackage[utf8]{inputenc}

\usepackage{amssymb,amsfonts,amsmath,bm}
\usepackage{graphics,graphicx}
\usepackage{xcolor}

\newcommand{\satya}[1]{#1}

\begin{document}

\title{\satya{One-dimensional Monte Carlo} dynamics at zero temperature}
\author{Alexei D. Chepelianskii}
\affiliation{Universit\'e Paris-Saclay, CNRS, Laboratoire de Physique des Solides, 91405, Orsay, France.}
\author{Satya N. Majumdar}
\affiliation{Universit\'e Paris-Saclay, CNRS, LPTMS, 91405, Orsay, France.}
\author{Hendrik Schawe}
\affiliation{LPTM, UMR 8089, CY Cergy Paris Universit\'e, CNRS, 95000 Cergy, France.}
\author{Emmanuel Trizac}
\affiliation{Universit\'e Paris-Saclay, CNRS, LPTMS, 91405, Orsay, France.}

\date{\today}

\begin{abstract}
We investigate, \satya{both analytically and with numerical simulations}, a Monte Carlo dynamics at zero temperature, where a random walker
evolving in continuous space and discrete time seeks to 
minimize its potential energy, by decreasing this quantity at each jump.
The resulting dynamics is universal in the sense that it 
does not depend on the underlying potential energy landscape,
as long as it admits a unique minimum; furthermore, the long time regime does not depend on the details of the jump distribution,
but only on its behaviour for small jumps. We work out the scaling properties of this dynamics, as embodied by the walker probability density. Our analytical predictions are in excellent agreement with direct 
Monte Carlo simulations.
\end{abstract}
\maketitle

\section{Introduction}
\label{sec:intro}

The steepest descent method is one of the oldest optimization schemes.
Cauchy, with a rather minimal mention in one of his papers, is credited
for its formulation \cite{Cauchy1847}.
It amounts to searching for the minimum of a well behaved function 
by following the steepest gradient ``downhill''. 
A stochastic reformulation has been proposed \cite{Robbins1951}, to alleviate the
computational cost of working in spaces with large dimensions:
the gradient is then estimated over a restricted set of directions,
drawn randomly \cite{Spall2012}. This technique is widely used in machine learning \cite{Sra_ML}.

We are interested here in a minimal version of stochastic gradient
descent, \satya{in a one-dimensional setting}. A random walker on the line, with position denoted by $x$, moves by random
jumps; the move is accepted only if it leads to the decrease of some
objective function $U(x)$, referred to as the potential.
\satya{In this respect, the walker is greedy, performing moves that always decrease $U$.}
We assume that $U$ admits a single minimum:
we are not interested in finding this point (taken as $x=0$ below),
but rather in the dynamics of the walker upon approaching it.
\satya{The specific form of the potential is immaterial; 
it does not need to be symmmetric, as long as it has no local minima,
beyond the global maximum at $x=0$.}
%it is symmetric ($U(x)=U(-x)$). 
%without (local) minima beyond the global minimum at $x=0$.
Such an algorithm can be viewed as the vanishing temperature limit
of a standard Metropolis sampling, a Monte-Carlo method where
a Markov chain is constructed to sample the phase space according to 
a predefined target distribution \cite{FrenkelSmith,NewmanBarkema,Krauth}.
As simple as it is --the walker position distribution function
is increasingly peaked at $x=0$--  such a dynamics exhibits non trivial features,
depending on the sampling chosen. A key role is indeed played by the 
probability distribution of attempted jumps, and in particular,
its behaviour for small jumps. 
\satya{The present $T=0$ problem differs from widespread 
approaches such as simulated annealing, where the temperature
ruling the evolution of a random walker is gradually decreased,
in order to find minima in a given potential landscape. Here, the motivation is different: the location of the 
potential minimum is known, and we are interested in the dynamics towards this target.}

Starting from the Metropolis rule, we define in section 
\ref{sec:model} the dynamics, in discrete time $n=0,1,\ldots$. The walker's density $P_n(x)$ evolves at long time towards
 $P_\infty(x) = \delta(x)$,
where $\delta$ denotes the Dirac distribution. 
Our goal is to resolve the approach
to this limiting form, that takes place in a self-similar way.
We show indeed in section 
\ref{sec:scaling} that $P_n(x)$ admits a scaling form at long times,
where the whole position and temporal information is encoded
in a single universal scaling function, independent of $U(x)$ and the initial condition. Its generic properties are
studied analytically. A number of exact 
solutions, presented in section \ref{sec:exact}
corroborate the general findings, and provide additional
insights into the long time dynamics. All predictions 
fare well compared to Monte Carlo simulations,
where the original dynamics is directly implemented.

\section{The Model}
\label{sec:model}

We consider a single particle moving on a line in the presence of 
an external confining potential $U(x)$, \satya{having a single minimum,
and no local minima. In other words, $U(x)$ should be a monotonically increasing function of $|x|$.}
At time $n$, an attempted jump $\eta_n$ is drawn from a distribution $w(\eta)$ and the particle moves according to the Metropolis rules
\begin{eqnarray}
x_n= \begin{cases}
x_{n-1} +\eta_n & {\rm with}\,\, {\rm prob.}\,\,\, 
p= {\rm min}\left(1, e^{-\beta\, \Delta U}\right)\quad
{\rm where} \quad \Delta U=U(x_{n-1}+\eta_n)-U(x_{n-1})  \\
\\
x_{n-1} & {\rm with}\,\, {\rm prob.}\,\,\, 1-p\, ,
\end{cases}
\label{Metropolis_T.1}
\end{eqnarray}
where $\beta$ is for inverse temperature $(k_B T)^{-1}$.
The position distribution $P_n(x)$ evolves via the Master equation
\begin{equation}
P_n(x) =  \int_{-\infty}^{\infty} dx'\, 
P_{n-1}(x')\, w(x-x')\, {\rm min}\left(1, e^{-\beta\, \left(U(x)-
U(x')\right)}\right) +
\left[1- \int_{-\infty}^{\infty} dx'\, w(x'-x)\, 
{\rm min}\left(1, e^{-\beta\, \left(U(x')-U(x)\right)}\right)\right]\, P_{n-1}(x),
\label{Master_T.1}
\end{equation}
where the jump distribution $w(\eta)$ is assumed to be 
symmetric: $w(\eta)=w(-\eta)$.
It is convenient to 
replace the `min' function above by the following identity
\begin{equation}
{\rm min}\left(1, e^{-\beta\, \left(U(x)-U(x')\right)}\right)=
\theta\left(U(x')-U(x)\right) +
e^{-\beta\, \left(U(x)-U(x')\right)}\, \theta\left(U(x)-U(x')\right)\,
\label{min_def}
\end{equation}
where $\theta(z)$ is the Heaviside theta function: $\theta(z)=1$ if $z>0$ and 
$\theta(z)=0$ if $z<0$. 

\begin{figure}
    \includegraphics[width=0.5\textwidth]{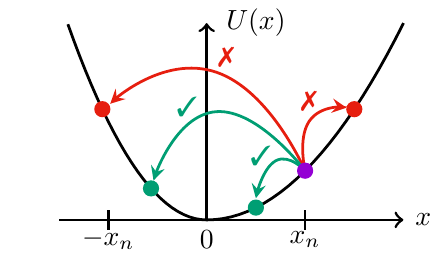}
    \caption{\label{fig1:zeroT}
        Possible moves of a particle at $T=0$ in the 
        potential $U(x)$, that monotonically increases with $|x|$.
        Only downhill stochastic moves are allowed \satya{(greedy motion)}.
        If at step $n$, the particle is at $x_n$, then at the next step
        it can jump only into the region $[-x_n,x_n]$. Any jump that takes the
        particle outside this interval is forbidden at $T=0$.
    } 
\end{figure}

We now focus on the limit $T=0$, i.e., $\beta=\infty$ \footnote{The $T=0$ limit is not innocuous, for it breaks detailed balance
\cite{Monthus}}. 
In this limit,
\eqref{min_def} becomes
\begin{equation}
{\rm min}\left(1, e^{-\beta\, \left(U(x)-U(x')\right)}\right)=
\theta\left(U(x')-U(x)\right)= \theta\left(|x'|-|x|\right)
\label{min_def_T0}
\end{equation}
where in arriving at the last equality we used the fact that the
potential increases monotonically with $|x|$ so that $U(x')>U(x)$
implies $|x'|> |x|$. Thus, in this limit, the particle can 
jump only downhill, and all uphill moves are forbidden.
More precisely, if the particle is at $x$ at step $n$, then
in the next step it can jump only to the region $x'\in [-|x|,|x|]$.
Any jump that takes it outside this region is forbidden at $T=0$ (see
Fig.~\ref{fig1:zeroT}).
Thus the explicit dependence of the position distribution
$P_n(x)$ on the form of the potential $U(x)$ drops out in
this $T=0$ limit, that can naturally be simulated by a Monte
Carlo method, as described in Appendix \ref{app:MC}.
However, even in this relatively simple limit, the  evolution
of the position distribution remains rather nontrivial. 
With the simplification in \eqref{min_def_T0}, the Master equation
\eqref{Master_T.1} reduces to a simpler form
\begin{equation}
P_n(x) =  \int_{-\infty}^{\infty} dx'\, 
P_{n-1}(x')\, w(x-x')\, \theta\left(|x'|-|x|\right)
+ \left[1- \int_{-\infty}^{\infty} dx'\, w(x'-x)\, \theta\left(|x|-|x'|\right)
\right]\, P_{n-1}(x) \, .
\label{Master2_T0}
\end{equation}
One can check that Eq.~(\ref{Master2_T0}) satisfies the probability 
conservation
\begin{equation}
\int_{-\infty}^{\infty} P_n(x)\, dx= \int_{-\infty}^{\infty} P_{n-1}(x)\, dx\, .
\label{prob_cons.1}
\end{equation}
Since the particle moves only downhill, we expect that at long times, i.e.,
in the limit $n\to \infty$, the position distribution should approach
a delta function at the origin (irrespective of the initial condition)
\begin{equation} 
\lim_{n\to \infty} P_n(x) = \delta(x)\, .
\label{p_stat.1}
\end{equation}
We are interested in computing how the position distribution relaxes
to this steady state.

\satya{For the analytical work, it is convenient} to start from a symmetric initial condition. This ensures
that at all times $P_n(x)$ is also symmetric, $P_n(x)=P_n(-x)$. In that case,
we can just focus on $x\ge 0$. 
\satya{This symmetry assumption will be tested against numerical
simulations, run in conditions of asymmetric initial conditions. It will be seen that the dynamics gradually suppresses non-symmetrical
features of the position distribution. An explicit solution 
in a specific case will confirm this computational observation}.
Our Master equation \eqref{Master2_T0} then reads (restricted to $x\ge 0$)
\begin{equation}
P_n(x)= \int_{-\infty}^{-x} dx'\, P_{n-1}(x')\, 
w(x-x') +  
\int_{x}^{\infty} dx'\, P_{n-1}(x')\,w(x-x')
+ \left[1- \int_{-x}^x dx'\, w(x'-x)\right]\, P_{n-1}(x)\, .
\label{Master3_T0}
\end{equation}
In the first integral on the right hand side (r.h.s) we make the change
of variable $x'\to -x'$ and use the symmetry $P_n(x)= P_{n}(-x)$  
to get for all $x\ge 0$
\begin{equation}
P_n(x)-P_{n-1}(x)= \int_{x}^{\infty} dx'\, P_{n-1}(x')\,
\left[ w(x+x')+w(x-x')\right]
-\left[\int_{-x}^x dx'\, w(x'-x)
\right]\, P_{n-1}(x)\, .
\label{Master4_T0}
\end{equation}

%%%%%%%%%%%%%%%%%%%%%%%%%%%%%%%%%%%%%%%%%%%%%%%%%%%%%%%%%%%%%%%
\section{Scaling analysis of the Master equation: asymptotic results}
\label{sec:scaling}

\subsection{The scaling ansatz}

The integral equation \eqref{Master4_T0} is still hard to solve 
exactly for all $n$ and for arbitrary symmetric jump distribution $w(\eta)$.
However, in the large $n$ limit, further simplifications occur.
Indeed, for large $n$ we make a scaling ansatz for $P_n(x)$ that should be
verified {\it a posteriori} and also confirmed numerically. Our ansatz reads
\begin{equation}
P_n(x) = B\, n^{\alpha}\, F\left( B\, n^{\alpha}\, x\right)\, ,
\label{scaling_ansatz.1}
\end{equation}
where the exponent $\alpha>0$ and the scale factor $B$ will be selected
subsequently. \satya{In doing so, we seek to ``resolve'' the structure of the asymptotic Dirac delta distribution, towards which the position distribution evolves}. 
Since $P_n(x)$ is taken here symmetric around $x=0$, the scaling function
$F(z)$ is symmetric: $F(z)=F(-z)$. In addition, from the normalization of $P_n(x)$, 
the scaling function $F(z)$
must satisfy the constraint
\begin{equation}
\int_{-\infty}^{\infty} F(z)\, dz=1\, .
\label{norm.0}
\end{equation}
The scaling form in \eqref{scaling_ansatz.1}  makes sense physically. It stipulates that the width
of the distribution decreases for large time $n$ as $\sim n^{-\alpha}$ with $\alpha>0$,
while the value at the peak increases as $n^{\alpha}$. Thus, as $n$ increases, the
position distribution function $P_n(x)$ gets more and more peaked near $x=0$, and eventually
approaches the delta function $P_n(x)\to \delta(x)$ as $n\to \infty$.

We substitute this scaling ansatz \eqref{scaling_ansatz.1} in the Master equation
\eqref{Master4_T0} and evaluate the left hand side (l.h.s) and the r.h.s.
For large $n$, the l.h.s simplifies to
\begin{equation}
P_n(x)-P_{n-1}(x)\approx \frac{\partial P_n(x)}{\partial n}\approx
B\,\alpha\, n^{\alpha-1}\,\left[F(z)+ z\, F'(z)\right]\, , \quad
{\rm where}\quad z= B\, n^{\alpha}\, x\, .
\label{scaling_lhs.1}
\end{equation}
We now substitute the scaling ansatz \eqref{scaling_ansatz.1} on the r.h.s
of \eqref{Master4_T0}. After making the change of variable
$z'= B\, n^{\alpha}\, x'$ inside the integrals on the r.h.s of
\eqref{Master4_T0}, it reads
\begin{equation}
\text{r.h.s} \approx \int_{z}^{\infty} dz'\, F(z')\, \left[ 
w\left(\frac{ z+z'}{B\, n^{\alpha}}\right) + 
w\left(\frac{z-z'}{B\, n^{\alpha}}\right)\right] 
-\left( \int_{-z}^{z} dz'\, w\left(\frac{|z'-z|}{B\, n^{\alpha}}\right)
\right)\, F(z)\, .
\label{rhs.1}
\end{equation}
Since we assumed $\alpha>0$ (to be verified {\it a posteriori}), 
it follows that as $n\to \infty$, the arguments of the function $w$ in \eqref{rhs.1}
approach $0$. Hence, the scaling behavior depends crucially on how the
jump distribution function $w(\eta)$ behaves for small $|\eta|$. We consider
the following natural class of power law behaviors near $\eta=0$
\begin{equation}
w(\eta) \approx C_p\, |\eta|^p \quad {\rm as} \quad \eta\to 0\, ,
\label{fx0.1}
\end{equation} 
where $C_p>0$ is a positive constant and $p>-1$ to ensure the normalization.
Substituting this behavior on the 
r.h.s in \eqref{rhs.1} and equating it to the l.h.s in
\eqref{scaling_lhs.1} we get, to leading order for large $n$,
\begin{equation}
B\,\alpha\, n^{\alpha-1}\,\left[F(z)+ z\, F'(z)\right]=
\frac{C_p}{B^{p}}\, n^{-\alpha p}\, \left[\int_z^{\infty} dz'\, \left[
(z+z')^p- (z'-z)^p\right]\, F(z')- \left(\int_{-z}^{z} dz'\, 
|z'-z|^p \right)\, F(z) \right]\, .
\label{lhs_rhs.1}
\end{equation} 
Note that $F'(z)= dF(z)/dz$.
In order that both sides scale as the same power of $n$ for large $n$, we
must have the exponent 
\begin{equation}
\alpha= \frac{1}{1+p}\, .
\label{alpha.1}
\end{equation}
Furthermore, let us choose the scale factor $B$ as
\begin{equation}
B= \left[C_p (1+p)\right]^{1/{1+p}}\, .
\label{B.1}
\end{equation}
With this choice of $B$, the scaling function $F(z)$ now depends only on the single parameter $p$, 
hence we will denote it by $F_p(z)$. It then satisfies the
integro-differential equation in $z\ge 0$, obtained from \eqref{lhs_rhs.1}
\begin{eqnarray}
F_p(z)+ z\, F_p'(z) &= & \int_z^{\infty} dz'\, \left[
(z'+z)^p+ (z'-z)^p\right]\, F_p(z')- \left(\int_{0}^{z} dz'\,
\left[(z+z')^p+ (z-z')^p\right] \right)\, F_p(z) \,  \nonumber \\
& = & \int_z^{\infty} dz'\, \left[
(z'+z)^p+ (z'-z)^p\right]\, F_p(z')- \frac{(2z)^{1+p}}{1+p}\, F_p(z)\, .
\label{Fpz_eqn.1}
\end{eqnarray}

To summarise, we justified {\it a posteriori} that the position distribution
$P_n(x)$ satisfies the scaling form
\begin{equation}
P_n(x)\approx B\, n^{1/(1+p)}\, F_p\left( B\, n^{1/(1+p)}\, x\right)\, , \quad {\rm where}\quad B= 
\left[C_p (1+p)\right]^{1/(1+p)}\, ,
\label{scaling_ansatz.2}
\end{equation}
and the scaling function $F_p(z)$, indexed by $p$, is determined from
the solution of the integro-differential equation \eqref{Fpz_eqn.1}. Note that
$F_p(z)$ is symmetric in $z$ and hence satisfies the
normalization condition
\begin{equation}
\int_0^{\infty} F_p(z)\, dz= \frac{1}{2}\, .
\label{norm.1}
\end{equation}
Thus the scaling function $F_p(z)$ depends only on the index $p$, but 
otherwise is universal, i.e., independent of the details of the jump 
distribution $w(\eta)$. For general $p>-1$, it is hard to solve the 
integro-differential \eqref{Fpz_eqn.1} exactly. However, as we show below,
one can derive the asymptotic behaviors of $F_p(z)$ as $z\to \infty$
and as $z\to 0$ for general $p>-1$. Later, we show that for the three
special cases, $p=0$, $p=1$ and $p=2$, it is possible to
obtain exact solutions for the full scaling function $F_p(z)$.

%%%%%%%%%%%%%%%%%%%
\subsection{Large $z$ behavior of $F_p(z)$}
For large $z$, the l.h.s of \eqref{Fpz_eqn.1} is dominated by the second term 
$z F_p'(z)$. In contrast, for large $z$, 
the first term on the r.h.s of \eqref{Fpz_eqn.1} goes to zero, 
while the second term dominates, as can be checked {\it a posteriori}. 
Hence, for large $z$, we get
$z\, F_p'(z) \approx - [(2z)^{p+1}/(p+1)]\, F_p(z)$. Integrating,
we obtain for large $z$ and any $p>-1$
\begin{equation}
F_p(z) \sim \exp\left[- \frac{{2}^{p+1}}{(1+p)^2}\, z^{1+p}\right]\, .
\label{Fpz_large.1}
\end{equation}
This result has a neat physical interpretation. In fact, expressing
the scaled variable $z= B\, n^{1/(p+1)} x$ in terms of the 
original distance variable $x$, one finds (up to power law prefactors)
\begin{equation}
P_n(x)\sim F_p(z) \sim 
\exp\left[- \frac{C_p}{1+p}\, (2\,x)^{p+1}
\, n\right]\, . 
\label{poisson.1}
\end{equation}
Hence $P_n(x)$ decays exponentially with increasing $n$ for fixed $x$. 
This can be understood
from a very simple Poisson type of argument. Consider the particle to be at $x$
at large time $n$
such that $B^{-1}\,  n^{-1/(1+p)}\ll |x|\ll 1$. In this regime 
$|z|=B\, n^{1/(1+p)} |x| \gg 1$. 
From this position
the particle attempts a jump. It succeeds if the jump takes it
into $x'\in [-x,x]$, otherwise it is unsuccessful and the particle stays at
$x$. Also, the influx of probability to $x$ from higher values
of $|x|$ is negligible in this regime.
Hence, the probability to stay at $x$ after step $n$ is
approximately 
\begin{equation}
P_n(x)\sim \left[1-p_{\rm accept}(x)\right]^n\, ,
\label{poisson.2}
\end{equation}
where the
acceptance probability at $x$ is given by
\begin{equation}
p_{\rm accept}(x)= \int_{-x}^{x} dx' w(x'-x)= \int_{0}^x \left[
w\left(x+x'\right)+ w\left(|x-x'|\right)\right]\, dx' . 
\label{p_accept.1}
\end{equation}
Since $|x|\ll 1$, the argument of 
$w$ is small and we can expand in Taylor series. 
Keeping only the leading order term for $|x|\ll 1$ and
using \eqref{fx0.1}, we obtain
\begin{equation}
p_{\rm accept}(x)\approx \frac{C_p}{(1+p)}\, (2\,x)^{1+p}\, .
\label{p_accept.2} 
\end{equation}
Substituting this result in \eqref{poisson.2}, we recover, for large $n$,
the result in \eqref{poisson.1}.

%%%%%%%%%%%%%%%%%%%%%%%%%%%%%%%%%%%%%%%%%%
\subsection{Small $z$ behavior of $F_p(z)$} 
Extraction 
of the small $z$ behavior of $F_p(z)$ from \eqref{Fpz_eqn.1}
is non trivial. We consider the r.h.s of \eqref{Fpz_eqn.1}
and start with the small $z$ behavior of the first term
$\int_z^{\infty} dz'\, \left[
(z'+z)^p+ (z'-z)^p\right]\, F_p(z')$. Consider the first integral
and rewrite it as
\begin{equation}
R_1=\int_z^{\infty} dz'\, (z'+z)^p\, F_p(z')= \int_{2z}^{\infty} du\, u^p\,
F_p(u-z)\, . 
\label{R1.1}
\end{equation}
We now expand $F_p(u-z)$ in Taylor series for small $z$ and keep terms up
to $O(z^2)$. Furthermore, we rewrite the integral $\int_{2z}^{\infty}=
\int_0^{\infty}- \int_0^{2z}$ and then expand the second part
also for small $z$. After some algebra, this gives for small $z$
\begin{equation}
R_1= b_0 + b_1\, z +b_2\, z^2 - \frac{F_p(0)}{1+p}\, (2z)^{p+1} + 
O(z^{\gamma})\, ,
\label{R1.2}
\end{equation}
with 
\begin{equation}
b_k= \frac{(-1)^k}{k!}\, \int_0^{\infty} du\, u^p\, F_p^{(k)}(u)\, ,
\quad{\rm where}\quad F_p^{(k)}(u)= \frac{dF_p^k(u)}{du^k} \, .
\label{bk.1}
\end{equation}
The exponent $\gamma$ is given by
\begin{equation}
\gamma= \min(3, p+2)\, .
\label{gamma.1}
\end{equation}
Repeating the same exercise with the second integral we get
\begin{eqnarray}
R_2 &= & \int_z^{\infty} dz'\, (z'-z)^p\, F_p(z')= \int_{0}^{\infty} du\, u^p\,
F_p(u+z) \nonumber \\
&=& b_0 + b_1\, z+ b_2\, z^2 + O(z^3)\, .
\label{R2.1}
\end{eqnarray}
Adding \eqref{R1.2} and \eqref{R2.1}, and substituting on the r.h.s
of Eq.~\eqref{Fpz_eqn.1}, we get
\begin{equation}
F_p(z)+ z\, F_p'(z)= \frac{d}{dz}[z F_p(z)]= 2\, b_0 + 2\, b_2\, z^2 - 2\frac{F_p(0)}{1+p}\, 
(2z)^{p+1} + O(z^{\gamma})
\label{Fpz_smallz.1}
\end{equation}
Integrating we get for small $z$
\begin{equation}
F_p(z) = 2\, b_0\,  + \frac{2 b_2}{3}\, z^2 - \frac{2^{p+2} F_p(0)}{(1+p)(2+p)}\, z^{1+p} + O(z^{\gamma})\, .
\label{Fpz_smallz.2}
\end{equation}
Thus, while the leading term is always $2 \, b_0$, the subleading term
depends on the value of $p$. Let us distinguish between three cases.

\begin{itemize}

\item {\bf {$p>1$:}} In this case, the first two terms in the small $z$
expansion of $F_p(z)$ are
\begin{equation}
F_p(z)\to a_0 + a_2\, z^2
\label{pgt1.1}
\end{equation}
where the two coefficients are given by
\begin{eqnarray}
a_0 &= & 2\, b_0 = 2 \int_0^{\infty} du\, u^p\, F_p(u) \label{a0.gt1}\\
a_2 &= & \frac{2 b_2}{3}= \frac{1}{3}\int_0^{\infty} du\, u^p\, F_p''(u)=
\frac{p(p-1)}{3}\, \int_0^{\infty} du\, u^{p-2}\, F_p(u)\, .\label{a2.gt1}
\end{eqnarray}
Note that since $a_2>0$, the function $F_p(z)$ for small $z$ actually 
increases as $z$ increases. Finally as $z\to \infty$, $F_p(z)$ has to decay
as in \eqref{Fpz_large.1}. Hence, the function $F_p(z)$
is non-monotonic as a function of $z$ for $p>1$. 
If one considers the symmetrized version of $F_p(z)$, 
there is a local minimum
(a hole) at $z=0$ (see Fig.~\ref{fig:p2} for the case $p=2$).

\item {\bf {$-1<p<1$:}} In this case, the first two terms are given by
\begin{equation}
F_p(z)\to a_0 + a_{p+1}\, z^{p+1}
\label{plt.1}
\end{equation}
where the coefficients read
\begin{eqnarray}
a_0 &= & 2\, b_0 = 2 \int_0^{\infty} du\, u^p\, F_p(u) \label{a0.lt1}\\
a_{p+1} & =& - \frac{2^{p+2} F_p(0)}{(1+p)(2+p)}\, . \label{ap.lt1}
\end{eqnarray}
Thus in this case, since $a_{p+1}<0$, the function $F_p(z)$ decreases
as $z$ increases from $0$, indicating that for $-1<p<1$, the scaling
function $F_p(z)$ is likely to
be a monotonically decreasing function of $z$, with a single peak at $z=0$,
see Fig.~\ref{fig:p0} for $p=0$. 

\item {\bf {$p=1$:}} Finally in the marginal case $p=1$, where
one has to merge the second and the third term together on
the r.h.s of \eqref{Fpz_smallz.2}, one gets
\begin{equation}
F_p(z)\to a_0 + a_{2}\, z^{2}
\label{peq.1}
\end{equation}
where the coefficients are
\begin{eqnarray}
a_0 &= & 2\, b_0 = 2 \int_0^{\infty} du\, u\, F_1(u) \label{a0.eq1}\\
a_{2} & =& \frac{1}{3}\, \int_0^{\infty} du\, u\, F_1''(u)- \frac{4}{3} F_1(0)
=-F_1(0) \, . \label{a2.eq1}
\end{eqnarray}
As we will see later, in this case we can derive the full solution exactly,
$F_1(z)=(1/\sqrt{\pi})\, e^{-z^2}$ whose small $z$ expansion agrees
with \eqref{a0.eq1} and \eqref{a2.eq1}. 

\end{itemize}

Let us then summarize the asymptotic behavior of $F_p(z)$ for general $p>-1$.
Using the symmetry of $F_p(z)$ we get
\begin{equation}
F_p(z)\sim \exp\left[- \frac{{2}^{p+1}}{(1+p)^2}\, |z|^{1+p}\right]\, \quad
{\rm as}\quad |z|\to \infty\, .
\label{Fpz_largez.3}
\end{equation}
The small $z$ behavior depends on the index $p$ and the first two terms 
are given by
\begin{eqnarray}
F_p(z) \to \begin{cases}
2\, \int_0^{\infty} du\, u^p\, F_p(u) - 
\left[\frac{2^{p+2}\, F_p(0)}{(1+p)(2+p)}\right]\, |z|^{p+1}\, 
& {\rm for}\quad -1<p<1 \\
\\
2\, \int_0^{\infty} du\, u^p\, F_1(u)- \left[F_1(0)\right]\, z^2 \, & {\rm for}\quad p=1 \\
\\
2\, \int_0^{\infty} du\, u^p\, F_p(u)+ 
\left[\frac{p(p-1)}{3} \int_0^{\infty} du\, u^{p-2}\, F_p(u)\right]\, z^2 \, & {\rm for}\quad p>1\, .
\end{cases}
\label{Fpz_smallz.3}  
\end{eqnarray}
There is thus a change of the small $z$ behavior at $p=1$,
from a decreasing function of $|z|$ for $|p|<1$ to an increasing
one for $p>1$. Indeed, with a probability density of jumps
$w(\eta) \propto |\eta|^p$, $p>1$ does penalize the occurrence
of the small jumps, required to move, as compared to
the case $p<1$. When $p>1$, the reduced likelihood of small jumps
leads to a depletion of $F_p(z)$ at $z=0$.
Beyond the asymptotic results 
derived here, we present in the next section exact solutions for the three special cases $p=0$, $p=1$
and $p=2$.

%%%%%%%%%%%%%%%%%%%%%%%%%%%%%%%%%%%%%%%%%%%%%%%%%%%%%%%%
\section{Exact scaling solutions}
\label{sec:exact}

\subsection{Exact solution for $p=0$}
\label{ssec:p0}

The case $p=0$ includes most natural symmetric jump distributions such as
\begin{itemize}

\item Gaussian: $w(\eta)= \frac{1}{\sqrt{2\,\pi}}\, e^{-\eta^2/2}$

\item Double-exponential: $w(\eta)= \frac{1}{2}\, e^{- |\eta|}$,
see also Appendix \ref{app:exact}.

\item Uniform distribution: $w(\eta)= \frac{1}{2}\,
\left[\theta(\eta+1)-\theta(\eta-1)\right]$.

\item Long-ranged distributions such as $w(\eta)= 1/[\pi (1+\eta^2)]$, the Cauchy jump distribution.

\end{itemize}
In all these cases, $w(\eta)\to {\rm const.}$ as $\eta\to 0$, implying $p=0$.
In this case, the integro-differential equation \eqref{Fpz_eqn.1} reads %(setting $p=0$)
\begin{equation}
F_0(z)+ z\, F_0'(z)= 2\, \int_z^{\infty} dz' F_0(z') - 2\, z\, F_0(z)\, ,
\label{F0z.1}
\end{equation}
and it must satisfy the normalization condition \eqref{norm.1}
together with the large $z$ asymptotic behavior \eqref{Fpz_largez.3}.
The solution of \eqref{F0z.1} turns out to be 
simple, as can be checked by direct substitution:
\begin{equation}
F_0(z)= e^{-2\, z}\, , \quad {\rm for}\quad z\ge 0.
\label{F0z.2}
\end{equation}
The large and small $z$ behaviors in \eqref{Fpz_largez.3}
and \eqref{Fpz_smallz.3} are consistent with the exact solution
\eqref{F0z.2}, as can be checked easily.
Thus for $p=0$, our prediction is that $P_n(x)$ is given by the scaling
form
\begin{equation}
P_n(x) \approx C_0\, n\, F_0\left( C_0\, n\, x\right)\, ,
\quad {\rm with}\quad F_0(z)= e^{-2 |z|}\, ,
\label{scalingp0.1}
\end{equation}
where we used the symmetry $F_0(z)=F_0(-z)$ and $C_0= w(0)$.
In Fig.~\ref{fig:p0}, we compare our theoretical prediction
for $F_0(z)$ with the numerical simulations (see Appendix
\ref{app:MC}
for details on the numerical aspects), obtained for three
different jump distributions all with $p=0$, namely the Gaussian,
the double-exponential and the uniform jump distributions. We find
excellent agreement between the theoretical prediction and the
simulations. \satya{The slight asymmetry observed between positive and negative values of $x$ is a fingerprint of the asymmetric initial
conditions. Its progressive disappearance gives an idea on the 
speed of convergence towards the scaling form. We come back to this question in Appendix \ref{app:exact}}.
 
\begin{figure}
    \includegraphics[width=0.48\textwidth]{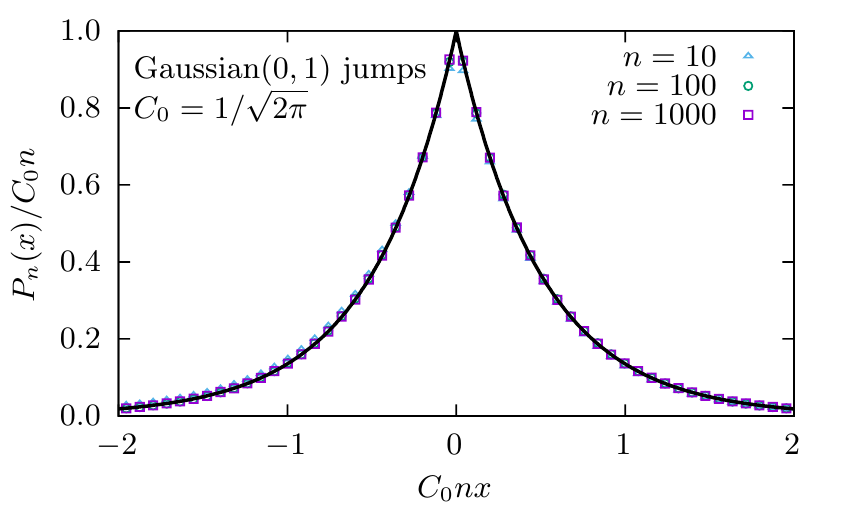}
    \includegraphics[width=0.48\textwidth]{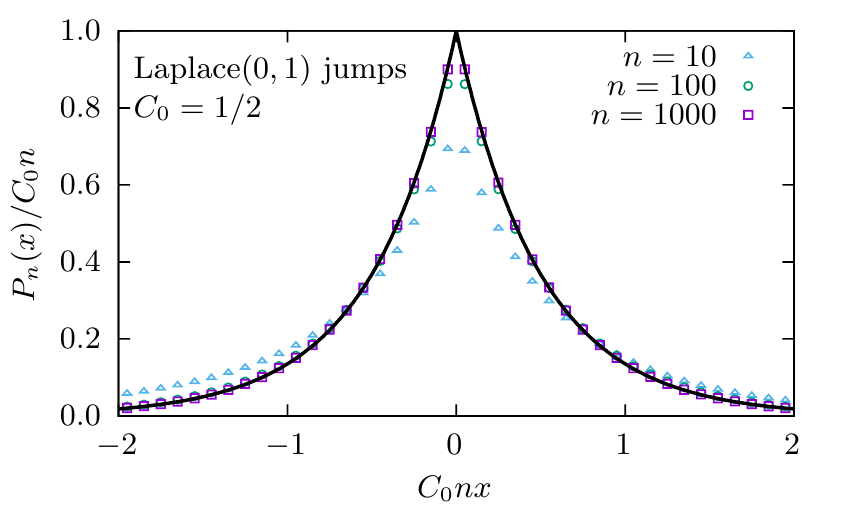}
    \includegraphics[width=0.48\textwidth]{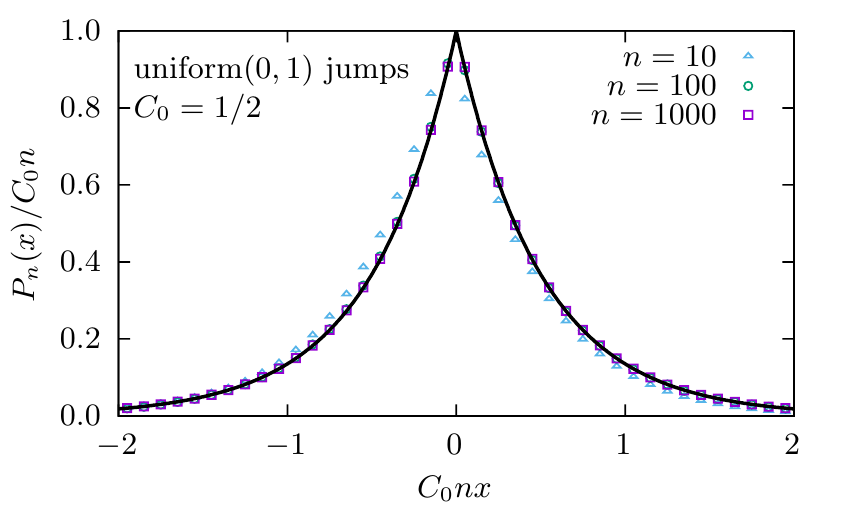}
    \includegraphics[width=0.48\textwidth]{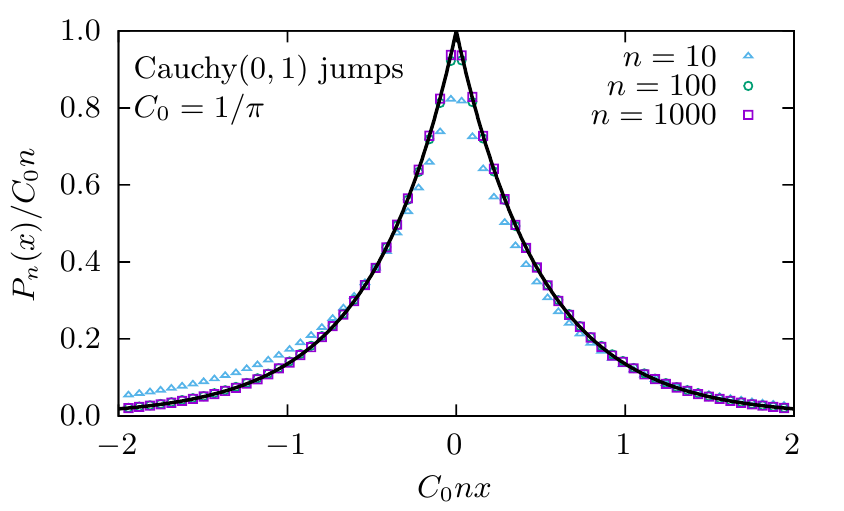}
    \caption{Case $p=0$.
        Scaling function
        $F_0(z)$ obtained from the simulation data and the scaled $P_n(x)$ as
        in \eqref{scalingp0.1}, compared with the theoretical 
        prediction of the universal form $F_0(z)=e^{-2|z|}$.  Four different
        jump distributions $w$ have been used: the Gaussian, the Laplace,
        the uniform and the Cauchy distributions, all having $p=0$.
        While the measurements for small times ($n=10$) show deviations 
        and an asymmetry due to the initial position chosen in the simulations ($x_0=-1$), the convergence 
        towards the form \eqref{scalingp0.1} is already very good for $n=1000$ in all 
        tested distributions. Measurements averaged over $10^7$ independent runs. 
    } 
    \label{fig:p0}
\end{figure}

\subsection{Exact solution for $p=1$} 
\begin{figure}
    \includegraphics[width=0.6\textwidth]{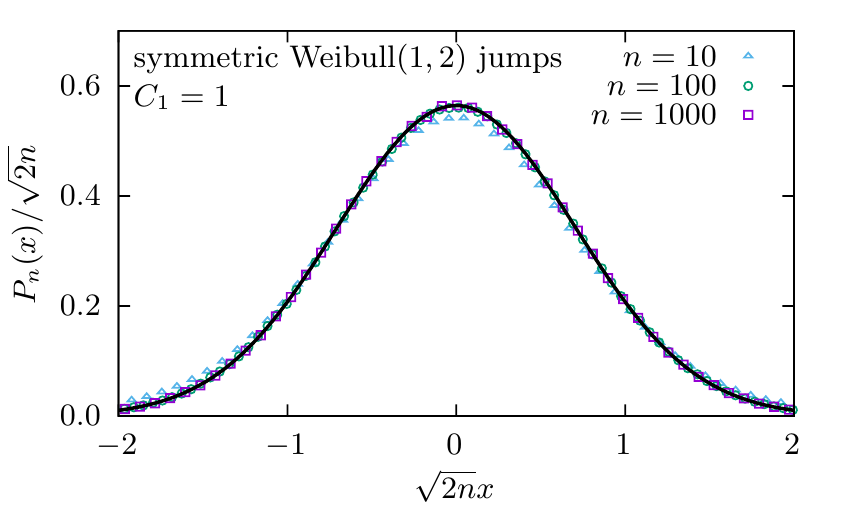}
    \caption{
        Case $p=1$. The scaling function
        $F_1(z)$ obtained from the simulation data for the scaled $P_n(x)$ as
        in \eqref{scalingp1.1}, compared with the theoretical 
        prediction of the universal form $F_1(z)=(1/\sqrt{\pi})\, e^{-z^2}$.
        In the simulation, we used the symmetric Weibull jump distribution 
        with shape factor 2 shown in \eqref{weibull.1}, thus $C_1=1$. 
        The agreement for $n \ge 100$ is excellent.
        Measurements obtained from $10^7$ independent runs 
        with $n=1000$ jumps each. \satya{Here again, the slight 
        $x\to -x$ asymmetry observed at early times is a consequence
        of having chosen an asymmetric initial condition.}
    }
    \label{fig:p1}
\end{figure}

An example of a 
jump distribution that belongs to the $p=1$ class is the symmetric Weibull
distribution (normalized to unity)
\begin{equation}
w(\eta)= |\eta|\, e^{-\eta^2}\, .
\label{weibull.1}
\end{equation}
In this case, setting $p=1$ in \eqref{Fpz_eqn.1} we get
\begin{equation}
F_1(z)+ z\, F_1'(z) = 2\, \int_z^{\infty} dz'\, z'\, F_1(z') - 
2\, z^2\, F_1(z)\, ,
\label{Fz1.1}
\end{equation}
where the solution $F_1(z)$ should satisfy the normalization condition
$\int_0^{\infty} F_1(z)\, dz=1/2$ and also the large $z$
asymptotic behavior in \eqref{Fpz_largez.3} with $p=1$.
Remarkably, \eqref{Fz1.1} also admits a simple solution (as can be 
checked by direct verification)
\begin{equation}
F_1(z)= \frac{1}{\sqrt{\pi}}\, e^{-z^2} \, ,
\label{Fz1.2}
\end{equation}
where the prefactor is chosen such that $\int_0^{\infty} F_1(z)\, dz=1/2$.
It is immediate to check that this exact solution is compatible
with the large and small $z$ behaviors in \eqref{Fpz_largez.3}
and \eqref{Fpz_smallz.3}. 
Thus, for $p=1$, our scaling prediction for the position distribution
$P_n(x)$ reads
\begin{equation}
P_n(x) \approx \sqrt{2\, C_1}\, n^{1/2}\, F_1\left( \sqrt{2\, C_1}\, n^{1/2}\, x\right)\, 
\quad {\rm with}\quad F_1(z)= \frac{1}{\sqrt{\pi}}\,e^{-z^2}\, ,
\label{scalingp1.1}
\end{equation}
where we again used the symmetry $F_1(z)=F_1(-z)$ and $B= \sqrt{2\, C_1}$ for $p=1$ 
from \eqref{scaling_ansatz.2}. In Fig.~\ref{fig:p1},
we compare the theoretical prediction for $F_1(z)$ in \eqref{scalingp1.1}
with the numerically obtained $F_1(z)$ using the symmetric Weibull
jump distribution in \eqref{weibull.1}, finding excellent agreement.

\subsection{Exact solution for $p=2$}

We now consider the case $p=2$. This corresponds to
symmetric jump distribution
$w(\eta)$ with the small $\eta$ behavior
\begin{equation}
w(\eta) \to C_2\, \eta^2 \quad {\rm as} \quad \eta\to 0\, .
\label{wsmall.1}
\end{equation}
Then, the scaling theory for the position distribution \eqref{scaling_ansatz.2}
predicts that for large $n$
\begin{equation}
P_n(x) \approx (3\, C_2\, n)^{1/3}\, F_2\left( (3\, C_2\,n)^{1/3}\, x\right)\, ,
\label{Pnx_scaling_p2.1}
\end{equation}
where the scaling function $F_2(z)=F_2(-z)$ is symmetric
and is normalized $\int_0^{\infty} F_2(z)\,dz=1/2$.
The scaling function $F_2(z)$, setting $p=2$ in \eqref{Fpz_eqn.1},
satisfies the integro-differential
equation (restricting only to $z\ge 0$)
\begin{equation}
F_2(z)+ z\, F_2'(z)= 2 z^2 \int_z^{\infty} dz'\, F_2(z') +
2\, \int_z^{\infty} dz'\, (z')^2\, F_2(z') - \frac{8 z^2}{3}\, F_2(z)\, ,
\label{diffeq_F2z.1}
\end{equation}
where $F_2'(z)= dF_2(z)/dz$. The function $F_2(z)$
must approach a constant as $z\to 0$ and should decay as
$F_2(z)\sim \exp[- 8 z^3/9]$ as $z\to \infty$: this follows from the general
asymptotics in \eqref{Fpz_largez.3} and \eqref{Fpz_smallz.3}.
In addition, it must satisfy the normalization condition \eqref{norm.1}.

To solve this integro-differential equation \eqref{diffeq_F2z.1}, 
the strategy would be first to reduce it to a differential equation.
To do this, let us define the cumulative scaling function
\begin{equation}
G_2(z)= \int_z^{\infty} F_2(z')\, dz'\, .
\label{cumul2.1}
\end{equation}
\satya{We have} $G_2'(z)=- F_2(z)$ and $G_2(z)\to 0$ as $z\to \infty$. Now consider
the second term on the r.h.s of \eqref{diffeq_F2z.1}. With an 
integration by parts, we can rewrite it as
\begin{equation}
-\int_z^{\infty} dz' (z')^2 G_2'(z)= z^2\, G_2(z) + 2\, \int_z^{\infty}
z'\, G_2(z')\, dz' \, .
\label{cumul2.2}
\end{equation}
Thus, using $F_2(z)=- G_2'(z)$, Eq.~(\ref{diffeq_F2z.1}) reads
\begin{equation}
-\left[1+\frac{8 z^3}{3}\right] G_2'(z)-z\, G_2''(z) - 4\, z^2\, G_2(z)= 4\, \int_z^{\infty} z'\, G_2(z') \, dz'\, .
\label{cumul2.3}
\end{equation}
Differentiating once more with respect to $z$ gives a differential equation
for $G_2(z)$
\begin{equation}
z\, G_2'''(z) + \left[2+\frac{8 z^3}{3}\right]\, G_2''(z) + 12 z^2 G_2'(z) + 4 z G_2(z)=0\, .
\label{cumul2.4}
\end{equation}
We divide by $z$, differentiate once more with respect to $z$ and use $G_2'(z)=- F_2(z)$
to finally obtain a third order ordinary differential equation for $F_2(z)$
\begin{equation}
F_2'''(z)+ \left[\frac{2}{z}+ \frac{8 z^2}{3}\right] F_2''(z)
+\left[- \frac{2}{z^2}+ \frac{52 z}{3}\right]\, F_2'(z)+ 16 F_2(z)=0\, .
\label{cumul2_final}
\end{equation}
A symbolic calculation software \cite{Mathematica} allows to solve this equation, and it gives the general
solution as a linear combination of three independent functions
%\begin{equation}
%F_2(z)= A_1\, {}_1F_1\left(\frac{1}{2},\frac{1}{3},-\frac{8 z^3}{9}\right)
%+A_2\, z^2\, {}_1F_1\left(\frac{7}{6}, \frac{5}{3}, - \frac{8 z^3}{9}\right)
%+ A_3\, {\rm MG}\left[\left\{\left\{-\frac{1}{3},\frac{1}{2}\right\}, 
%\left\{ \right\} \right\},\left\{\left\{-\frac{1}{3},-\frac{2}{3}\right\},\left\{ \right\}\right\}, -\frac{8 z^3}{9}\right]\, ,
%\label{F2z_sol.1}
%\end{equation} 
\begin{equation}
F_2(z)= (A_1)\, {}_1F_1\left(\frac{1}{2},\frac{1}{3},-\frac{8 z^3}{9}\right)
+(A_2)\, z^2\, {}_1F_1\left(\frac{7}{6}, \frac{5}{3}, - \frac{8 z^3}{9}\right)
+ (A_3)\,G_{2,2}^{2,2}\left(
\begin{array}{l}
-\frac{1}{3}, \phantom{-}\frac{1}{2}\\
-\frac{1}{3}, -\frac{2}{3}
\end{array}\middle\vert
-\frac{8z^3}{9}
\right)\, ,
\label{F2z_sol.1}
\end{equation} 
where ${}_1F_1(a,b,z)$ is the Kummer's confluent hypergeometric function
and $G_{m,n}^{p,q}$ denotes the Meijer's G function. The constants $A_1$, $A_2$ and $A_2$
have to be determined from boundray conditions.
First we note that the third function $G_{m,n}^{p,q}$ diverges as $1/z$ as $z\to 0$.
Since this is not allowed (the scaling function $F_2(z)$
is normalizable and approaches a constant as $z\to 0$ from
\eqref{Fpz_smallz.3}). Hence we must have $A_3=0$. Thus
\begin{equation}
F_2(z)= A_1\, \left[{}_1F_1\left(\frac{1}{2},\frac{1}{3},-\frac{8 z^3}{9}\right)
+ B_2\, z^2\, {}_1F_1\left(\frac{7}{6}, \frac{5}{3}, - 
\frac{8 z^3}{9}\right)\right]\, ,
\label{f2z_sol.2}
\end{equation}
where we define $B_2= A_2/A_1$. The overall global constant $A_1$ can be fixed by the normalization condition in \eqref{norm.1}.
The only remaining unknown constant $B_2$ has to be found from
the boundary condition as $z\to \infty$, where we expect
from \eqref{Fpz_largez.3} with $p=2$ that
$F_2(z) \sim e^{- 8 z^3/9}$. So, the constant $B_2$ has to be chosen such that
$F_2(z)$ decays in this fashion as $z\to \infty$.

\begin{figure}
    \includegraphics[width=0.6\textwidth]{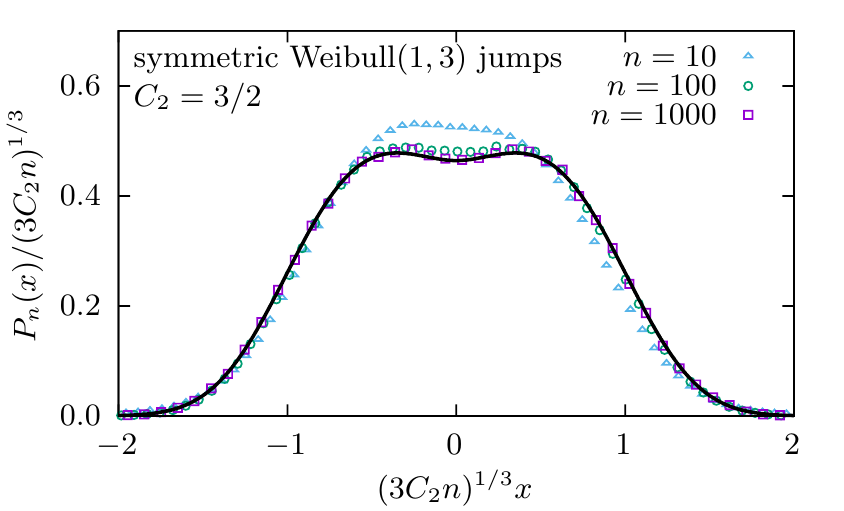}
    \caption{
        Case $p=2$. Comparison between the numerically obtained scaling function
        $F_2(z)$ and the prediction of Eq.~\eqref{sol_p2.1}.
        The scaling function is symmetric ($F_2(-z)=F_2(z)$), non-monotonic, with a pair of maxima at 
        $z=\pm 0.350322\dots$ and a local minimum at $z=0$.
        The simulation data for $P_n(x)$, scaled as
        in \eqref{scaling_ansatz.1}, are shown with the symbols;
        they have been obtained with a normalized jump distribution of the symmetric Weibull type
        and shape factor $3$, i.e.,
        $w(\eta)= \frac{3}{2}\, \eta^2\, e^{- |\eta|^3}$, thus $C_2=3/2$. 
        The agreement for $n = 1000$ is excellent. Measurements obtained from $10^7$ independent runs 
        with $n=1000$ jumps each. \satya{Here again, the asymmetry 
        that is visible at $n=10$ stems from our initial condition. Asymmetry is gradually washed out as time proceeds}.
    }
    \label{fig:p2}
\end{figure}

To fix $B_2$, we need to find the asymptotic
behavior of ${}_1F_1(a,b,-x)$ when $x$ approaches
$\infty$ along the real line and $\theta=a-b$ is
a non-integer.
It turns out that this asymptotic behavior 
is rather subtle and has been obtained only rather 
recently~\cite{Paris2013}. For large $x$ one gets
\begin{equation}
{}_1F_1(a,b,-x)= \frac{x^{-a}\, \Gamma(b)}{\Gamma(b-a)}\sum_{k=0}^{\infty} 
\frac{(a)_k\, (1+a-b)_k}{k!\, x^k} + x^{a-b}\, e^{-x} \left[ \cos(\pi(a-b))\
 + O(1/x)\right]\, ,
\label{asymp_ch.1}
\end{equation}
where $(a)_k= \Gamma(a+k)/\Gamma(a)$ is the Pochamer symbol. 
Thus, generically, to leading order, it decays as a 
power law $\sim x^{-a}$ as $x\to \infty$. 
Substituting this asymptotic behavior in \eqref{f2z_sol.2} using
$x= 8 z^3/9$, we get as $z\to \infty$ 
\begin{eqnarray}
F_2(z)&\approx & A_1\, \left[ \frac{\Gamma(5/3)}{\Gamma(1/2)} 
\left(-\frac{8}{9}\right)^{-7/6}\left(B_2 - \frac{\pi}{3 \Gamma^2(5/6)}\,
 \left(\frac{2}{3}\right)^{1/3}\right)\, z^{-3/2}\,\sum_{k=0}^{\infty} 
\frac{9^k \,\Gamma(1/2+k) \Gamma(7/6+k)}{k!\,8^k\, z^{3k}}\right]  \nonumber \\
&+ & A_1\,  \frac{\Gamma(1/3)\cos(\pi/6)}{\sqrt{\pi}}\,\left(\frac{8}{9}\right)^{1/6}\,
 z^{1/2}\, e^{-8 z^3/9} \, .
\label{f2z_asymp.2}
\end{eqnarray}
Since the boundary condition for large $z$ predicts that
$F_2(z)\sim \exp[- 8 z^3/9]$ (see Eq.~(\ref{Fpz_largez.3})),
we must eliminate the slow power law decay in \eqref{f2z_asymp.2}.
This can be done by choosing the constant $B_2$ as
\begin{equation}
B_2= \frac{\pi}{3 \Gamma^2(5/6)}\,
 \left(\frac{2}{3}\right)^{1/3} \, .
\label{B2.1}
\end{equation}
Finally, the global constant $A_1$ is obtained from the normalization condition
$\int_0^{\infty} F_2(z)\, dz=1/2$. This gives (upon using Mathematica to do the
integral)
\begin{equation}
A_1= \left(\frac{3}{2}\right)^{1/3} \frac{\Gamma^2(5/6)}{\pi}\, .
\label{A_1.1}
\end{equation}
Thus, with the two unknown constants $A_1$ and $B_2$ fixed, we then
have our exact scaling function valid for all $z\ge 0$ (and symmetrically
for $z\le 0$)
\begin{equation}
F_2(z)= \left(\frac{3}{2}\right)^{1/3} \frac{\Gamma^2(5/6)}{\pi}\,
{}_1F_1\left(\frac{1}{2},\frac{1}{3},-\frac{8 z^3}{9}\right)
+ \frac{z^2}{3}\, 
{}_1F_1\left(\frac{7}{6}, \frac{5}{3}, - \frac{8 z^3}{9}\right)\, .
\label{sol_p2.1}
\end{equation}
The function $F_2(z)$ has the following asymptotic
behaviors
\begin{eqnarray}
F_2(z) \approx \begin{cases}
\left(\frac{3}{2}\right)^{1/3} \frac{\Gamma^2(5/6)}{\pi} + \frac{1}{3}\, z^2
+O(z^3) & {\rm as}\,\, z\to 0 \\
\\
b\, \sqrt{z}\, e^{- 8 z^3/9} & {\rm as} \,\, z\to \infty
\end{cases}
\label{F2z_asymp}
\end{eqnarray}
where the constant prefactor $b$ is given by
\begin{equation}
b= \frac{\sqrt{3}}{2^{5/6}\, \pi^{3/2}}\, \Gamma^2(5/6)\, \Gamma(1/3)= 0.595887\ldots
\label{b_exact.1}
\end{equation}
The large $z$ asymptotic follows from the remaining nonzero term
in \eqref{f2z_asymp.2} once $B_2$ is fixed. One can check that
the small $z$ behavior above is also completely in agreement
with \eqref{Fpz_smallz.3}.
A plot of this function $F_2(z)$ vs.\ $z$ is given in Fig.~\ref{fig:p2}.
In this Figure, we also compare our theoretical prediction
with numerical simulations for $p=2$, finding excellent agreement.
Interestingly, the function $F_2(z)$
is non-monotonic, and has a pair of maxima at $z\approx \pm 0.350322\dots$. 
The reason behind this non monotonicity has been put forward below Eq.~\eqref{Fpz_smallz.3}: $p>1$ penalizes small jumps.
Here, we consequently expect the dynamics to proceed
in back and forth motion, from the left of the minimum to the right,
and vice-versa for the subsequent successful jump. Such an oscillatory motion is fully compatible with the ``two hump'' structure 
of the scaling function for $p>1$. Yet, we stress that
this oscillatory motion towards $x=0$ is not specific to cases with $p>1$:
the late-time probability that the position changes sign after an accepted move is smoothly increasing when $p$ increases.
This probability is given by 
\begin{equation}
    \frac{\int_x^{2x} \eta^p \, d\eta  }{\int_0^{2x} \eta^p \, d\eta } \,=\,
    1-\frac{1}{2^{p+1}} .
\end{equation}
Starting from 0 when $p \to -1^+$, it thus crosses 50\% for $p=0$, and exceeds 90\% for
for $p>2.33$.

\begin{figure}
    \includegraphics[scale=1]{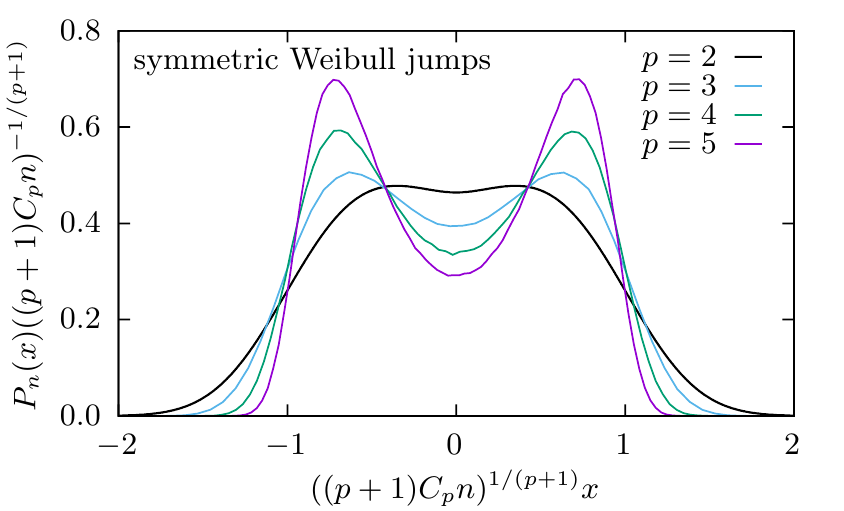}
    \caption{
        Numerically obtained asymptotic scaling functions $F_p(z)$ 
        vs the scaling variable $z$ for $p=3, 4, 5$.
        The Monte Carlo simulations were run with a normalized jump distribution of the symmetric Weibull type
        and shape factor $(p+1)$, i.e.,
        $w(\eta)= \frac{p+1}{2}\, \eta^p\, e^{- |\eta|^{p+1}}$.
        Measurements are collected over $10^6$ independent runs 
        with $n=10^5$ jumps for $p=3, 4$ and $n=10^6$ jumps for $p=5$.
        The curve for $p=2$ is Eq.~\eqref{sol_p2.1} shown for comparison.
    }
    \label{fig:p_higher}
\end{figure}

Finally, we show in Fig.~\ref{fig:p_higher} the evolution of the scaling
function $F_p$ with $p$, as obtained in Monte Carlo simulations. 
The two hump structure 
resulting from increasing $p$ is visible. It is a consequence of the reduced 
likelihood of performing small jumps, when $p$ increases, as discussed at the end of section 
\ref{sec:scaling}.

\section{Concluding remarks}

We have studied a stochastic steepest-gradient descent on a line,
in a potential landscape $U(x)$. A random walker proceeds in discrete time, with a succession of jumps ($\eta_n$ at time $n$).
\satya{The walker is greedy, in the sense that it only performs
moves that decrease the energy $U(x)$.}
The dynamics does not depend 
on $U(x)$, and is driven to its minimum at $x=0$, irrespective
of the value (or even the existence) of the gradient,
as long as the minimum is non-degenerate \satya{(no local minima)}. 
The long-time regime has been shown to be self-similar,
with a scaling function $F_p(z)$ that only depends on the 
likelihood of small jump displacements, through the 
parameter $p$: Denoting the jump probability distribution
by $w(\eta)$, we have $w(\eta) \propto \eta^p$ for $\eta\to 0$
(and thus $p=0$ when $w(0)$ is finite). 
A Poisson-type of argument reveals that the large $z$ tail
of the scaling function is of the form 
$\log F_p(z) \propto -z^{1+p}$, and it also provides
the scaling variable as $z=x\,n^{1/(1+p)}$. 
In other words, the root mean squared
(r.m.s.) spread of the particle's 
position shrinks at large times $n$ as $n^{-1/(1+p)}$.
It is quite natural that large $p$ values lead to a dynamical 
slow down, since they are associated to a smaller
likelihood of small jumps, which is detrimental to
evolution. A byproduct of the analysis is that 
redefining the clock from $n$ to $\cal N$, where $\cal N$ would count the number of accepted moves, the scaling variable would become
$z= {\cal N} x$, while the form of the scaling function 
$F_p$ would of course be unaffected.

\begin{figure}[htb]
    \includegraphics[scale=1]{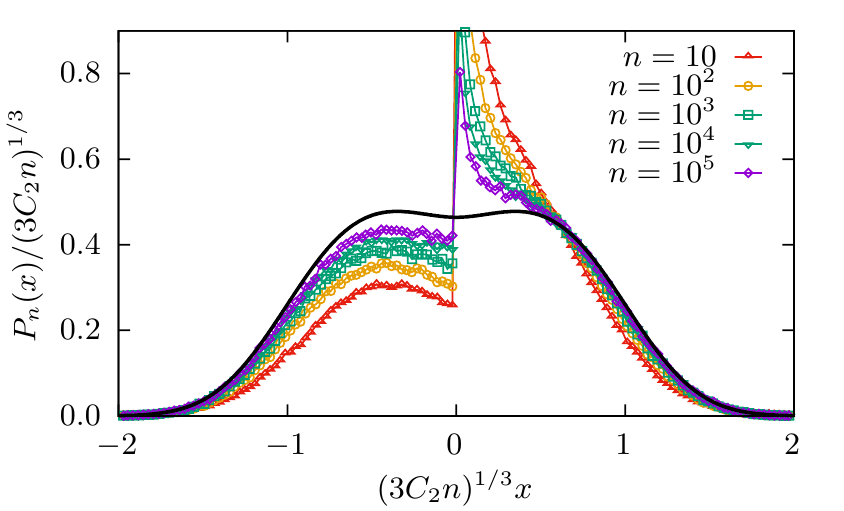}
    \includegraphics[scale=1]{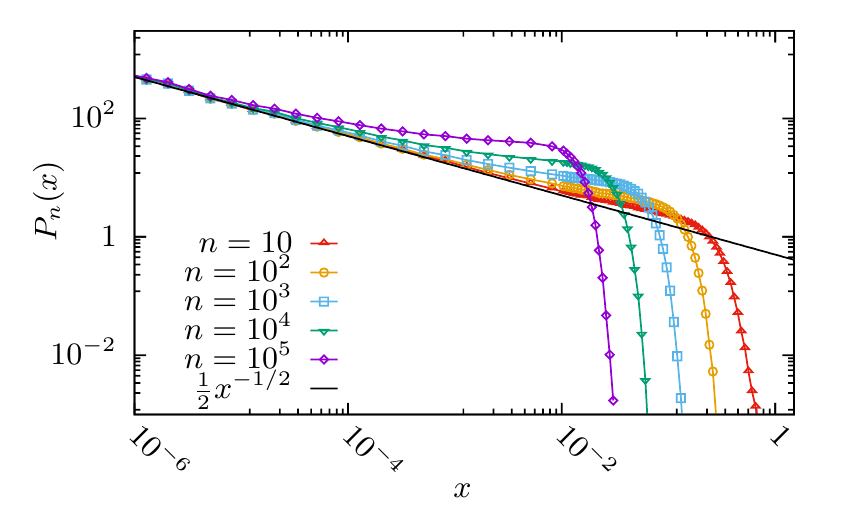}
    \caption{
    Monte Carlo results for initial conditions with a singularity 
        at $x=0$, here $P_0(x) = \frac{1}{2} x^{-1/2}$
        for $0<x<1$, and $P_0(x)=0$ otherwise;
        the jumps are of the symmetric Weibull type ($p=2$).
        The left panel shows the measured $P_n(x)$ after $n$ 
        jumps averaged over $10^7$ realizations and 
        scaled to demonstrate the slow convergence towards 
        the scaling form $F_2(z)$ shown as a black line.
        \satya{The right panel shows $P_n(x)$ on a log-log scale,
        as a function of the unscaled variable $x$. Comparison of the two panels shows that close to $x=0$, 
        $P_n(x)$ is of the form $x^{-1/2}$, before it crosses over to the asymptotic shape, further away from $x=0$}.
    }
    \label{fig:zero_x}
\end{figure}

We restricted the analysis to
the search of the large-time symmetric scaling solution, which is universal in the sense that it is independent of initial conditions. The excellent agreement with the Monte Carlo simulations, 
that start from a non symmetric initial configuration, proves that asymmetric modes
must decay faster than the symmetric ones.
In some situations, the decay is slow. Such is the case when 
the initial distribution 
$P_0(x)$ is of the type $x^{-\mu}$ for small $x$ with $0<\mu<1$, thus singular near the origin. 
By construction, the smaller an $x$-value, the 
least probable it will be affected by the dynamics. We then 
expect that in such a case, $P_n(x)$ features the same
singularity as $P_0(x)$ close enough to $x=0$. This is confirmed 
by Fig.~\ref{fig:zero_x}, where $\mu=1/2$. The initial asymmetry 
of $P_0(x)$ (\satya{note that $P_0(x)=0$ for $x<0$}), impinges on later times,
while the singularity at $x=0$ has a rather spectacular 
effect on $P_n(x)$. Yet, Fig.~\ref{fig:zero_x} gives credit to the statement that 
for $n\ \to \infty$, the scaled $P_n$ tends towards 
$F_2$, as given in Eq.~\eqref{f2z_sol.2}.
In addition, explicit exact calculations for the double exponential
jump distribution, beyond scaling, show  
1) the convergence towards the symmetric scaling  form
and 2) the faster decay of asymmetric contributions
coming from the initial conditions. Details 
are provided in Appendix \ref{app:exact}.

%However, proving this is presumably a difficult task, because it would require to compute the non universal (initial condition dependent) corrections to scaling. 

We did not discuss the cases where $w(\eta)$ is
depleted near $\eta=0$, with a vanishing probability
in some interval $[-\eta^*,\eta^*]$: such a case 
would be non self-averaging, and
lead to a dynamical arrest, whenever the walker
arrives within the depletion segment $[-\eta^*,\eta^*]$.
One would need to average over may realizations
sharing the same initial conditions in order to 
obtain an interesting a smooth late-time dynamics,
where $P_n(x) \to \delta(x)$, thereby resolving the structure 
of the $\delta$ peak. Conversely, in the case we have studied,
and although we have averaged our Monte Carlo data over 
a large number of samples to garner statistics,
the dynamics is self-averaging in the sense that a single 
trajectory leads to a well defined scaling function; 
statistics is increased by following the
evolution on longer time scales.

In this paper, we focused on $T=0$ where the steady state is trivial, but the late time relaxational dynamics is self-similar and typical observables (such as the r.m.s displacement) decay as a power law in time $n$ for large $n$. This power law behavior is due to the vanishing acceptance probability of new moves at the minimum.  
 In a recent paper~\cite{CMST1}, we studied the finite temperature version of this model
 where the dynamics satisfies detailed balance. In this case the steady state is of Gibbs-Boltzmann form and the relaxation of observables towards their steady state value becomes exponential with the error due to incomplete convergence decaying as  $\Lambda^n$ (with $\Lambda<1$). The relaxation time $\tau=-1/\ln \Lambda$ has a rich behavior as a function of the jump amplitude $a$, achieving a minimum value at an optimal $a^*$. Surprisingly for $a > a^*$, the relaxation at finite temperature is governed by self-similar scaling solutions very similar to the scaling ansatz established here for $T=0$. The framework developed here may thus find applications beyond the $T=0$ limit. 

\begin{appendix}
\section{Monte Carlo simulations}
\label{app:MC}

The zero temperature dynamics investigated here
defines a Markov chain, that is naturally simulated
with the Monte Carlo technique \cite{FrenkelSmith,NewmanBarkema,Krauth}.
In order to observe the relaxation of the particle distribution $P_n(x)$
we directly simulate $m$ independent particles (here $m = 10^6$ or $m = 10^7$), all starting at $x_0 = -1$
by iterating
\begin{align}
    x_n= \begin{cases}
        x_{n-1} + \eta_n & {\rm if}\,\, \left| x_{n-1} + \eta_n \right| < \left| x_{n-1} \right|\\
        x_{n-1} & {\rm otherwise} ,
        \end{cases},
\end{align}
which is a simplified but equivalent version of Eq.~\eqref{Metropolis_T.1} using Eq.~\eqref{min_def_T0}.
The $\eta_n$ are independent random numbers drawn from the respective 
distributions $w(\eta)$, which
(except for the Gaussian for which we used an implementation of the Ziggurat method \cite{marsaglia2000ziggurat})
can be generated using the inversion method \cite{press2007numerical}.
For each value of $n$ in which we are interested, e.g., $n=10, 100, 1000$,
we initialize a histogram and update it with the position of the $m$ particles, after
iteration $n$. We use the shape of the distribution we measured this way at different values of $n$
to determine whether our Markov chain is long enough to reach the predicted scaling form, 
which typically happens very fast for jump distributions with low values of $p$ and takes a longer time for
jump distributions with larger values of $p$.

\section{An exact solution for the double-exponential jump distribution}
\label{app:exact}

To find a case where the long time asymptotic properties of the probability distribution $P_n(x)$ can be obtained exactly, we use the $p=0$ type
jump distribution:
\begin{align}
w(\eta) = \frac{1}{2} e^{-|\eta|}.
\end{align}
%We assume a symmetric potential, $U(x) = U(-x)$, strictly increasing for $x > 0$ ($U(x') > U(x)$ for $x' > x > 0$).

It is convenient to separate $P_n(x)$ into symmetric and anti-symmetric components: 
\begin{align}
P_n(x) = S_n(x) + Z_n(x)  
\end{align}
with $S_n(x) = S_n(-x)$ and $Z_n(x) = -Z_n(-x)$. 
Substituting this decomposition in the Master equation
(\ref{Master2_T0}) and \satya{using the symmetry of
%of the potential $U(x)$ and 
$w(\eta)$}, one finds that the associated Master equations for $S_n$ and $Z_n$ separate. We first focus on the symmetric component $S_n(x)$. For $x> 0$, the corresponding Master equation reads:
\begin{align}
S_{n+1}(x) = \cosh(x) \int_{x}^{\infty} e^{-y} S_n(y) dy + R(x) S_n(x)
\end{align}
where we introduced a compact notation for the probability to reject an attempted move from $x$:
\begin{align}
R(x) = \int_{-\infty}^{\infty} dy\, w(y-x)\, \theta\left(|y|-|x|\right) = e^{-x} \cosh x.
\end{align}
This quantity ranges from 1 at small $x$ (where most moves are rejected), to 1/2 at large $x$, where downhill moves only are accepted.

Introducing the generating function:
\begin{align}
Q_s(x) = \sum_{n\ge 0} s^n S_n(x) ,
\end{align}
we find the equation: 
\begin{align}
\cosh(x) \int_{x}^{\infty} e^{-y} Q_s(y) dy + (e^{-x} \cosh x) Q_s(x) = \frac{Q_s(x) - S_0(x)}{s} .
\end{align} 
This integral equation can be reduced to a first order differential equation, that can be integrated explicitly:
\begin{align}
Q_s(x) = \frac{S_0(x)}{1 - s e^{-x} \cosh x} + 2 s (e^{2 x} + e^{4 x}) h_s(x)^{-1-\frac{1}{2-s}} \int_x^{\infty} dy \; h_s(y)^{-1 + \frac{1}{2 - s}} S_0(y) ,
\label{eqQs}
  \end{align}
where we introduced the notation:
\begin{align}
  h_s(x) &= e^{2 x}(2 - s) - s .
\end{align}  
With this expression, we can use the general technique of singularity analysis \cite{acombinatorics} to obtain exact results on the large $n$ behavior of $S_n(x)$, scrutinizing the complex plane position of the singularities of the generating function $Q_s(x)$ as function of the parameter $s$.

To make further progress, we assume $S_0(x) = \delta(|x| - y)/2$ (with $y > 0$). This choice is the symmetric part of $P_0(x) = \delta(x - y)$ and we find: 
\begin{align}
2 Q_s(x) = \frac{\delta(x - y)}{1 - s e^{-y} \cosh y} + 2 s (e^{2 x} + e^{4 x}) h_s(x)^{-1-\frac{1}{2-s}} \; h_s(y)^{-1 + \frac{1}{2 - s}} \theta(y - x) .
  \end{align}
Singularities appear when $h_s(x) = 0$ or $h_s(y) = 0$. For $x < y$, the singularity in the complex $s$-plane that is closest to the origin $|s| = 0$ is the solution of $h_s(x) = 0$, and reads
\begin{align}
s = \frac{2}{1 + e^{-2 x}} = R(x)^{-1} . % \frac{1}{e^{-x} \cosh x}
\end{align}
This implies that up to sub-exponential factors:
\begin{align}
    S_n(x) \sim R(x)^{n}.
\end{align}
Particles cannot climb uphill in this model and $R(x') < R(x)$ for $x' > x > 0$. It is thus not surprising that the decay rate of $S_n(x)$ is determined by the rejection probability $R(x)$.

\begin{figure}[htb]
    \includegraphics[scale=1]{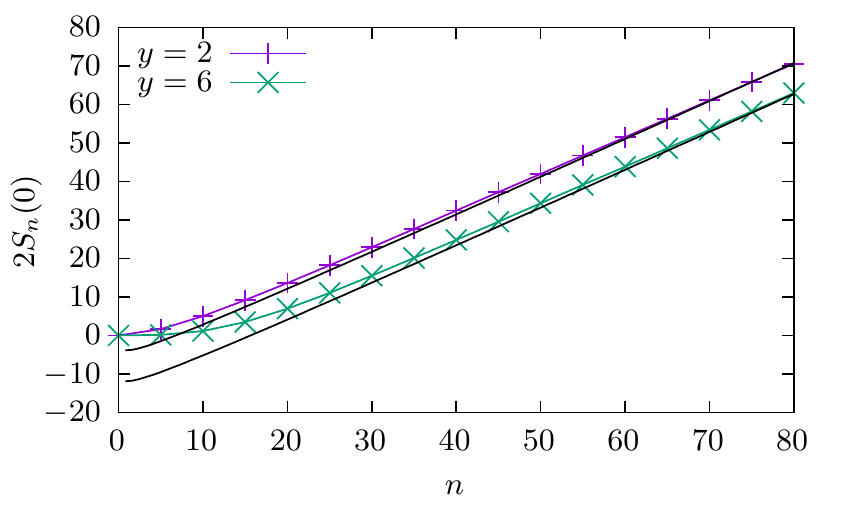}
    \caption{Comparison between the asymptotic behavior $2 S_n(0)$ and numerical simulations for $S_0(x) = \delta(|x|-y)/2$, with $y = 2$ and $y = 6$. Symbols show numerical results and continuous black lines show Eq.~(\ref{eqP0zeroTemp}).
    }
    \label{fig:theoexactvsnum}
\end{figure}

It is possible to obtain sub-exponential corrections for particular values of $x$.
For example for $x = 0$, the singularity with the smallest modulus is at $s = 1$. Expanding the generating function around this position, we find:
\begin{align}
2 Q_s(0) \simeq \frac{1}{(1-s)^2} - \frac{1}{1-s}\log\frac{1}{1-s} - \frac{1+\log(e^{2y}-1)/2}{1-s} + ... 
\end{align}
This allows us to obtain the sub-exponential corrections up to the first term that depends on the initial position $y$: 
\begin{align}
2 S_n(0) \simeq n - \log n - \left(1+ \gamma -\log 2 + \log[e^{2 y}-1] \right) + ... ,
\label{eqP0zeroTemp}
\end{align}
where $\gamma$ is the Euler-Mascheroni constant. This formula is in good agreement with numerical simulations which are shown shown in Fig.~(\ref{fig:theoexactvsnum}) for two different initial positions $y$ of the random walker.

A similar asymptotic analysis for $x \ll 1$ gives: 
\begin{align}
2 S_n(x) \simeq n^{1+x} R(x)^n + ...,
\end{align}
where we derived only the leading sub-exponential term. 
Setting ${\widetilde x} = n x$, we can get:
%we can check that we indeed recover the scaling function ansatz: 
\begin{align}
S_n(n^{-1} {\widetilde x}) \, \simeq \, \frac{n}{2} \left(1 - \frac{|{\widetilde x}|}{n}\right)^n \, \simeq \, \frac{n}{2}\, e^{-|{\widetilde x}|} .
\end{align}
Using the result $F_0(z) = e^{-2z}$ derived in section \ref{ssec:p0} for the present 
$p=0$ situation, together with the fact that here, $C_0=B=1/2$,
we recover the limiting scaling behaviour derived
in section \ref{ssec:p0}: 
$P_n(x) = n F_0(nx/2) /2$, see Eq. \eqref{scalingp0.1}.

For the anti-symmetric component of the probability distribution $Z_n(x)$, the Master equation for $x>0$ is:
\begin{align}
 &Z_{n+1}(x) = \sinh(x) \int_{x}^{\infty} e^{-y} Z_n(y) dy + R(x) Z_n(x) .
 \end{align}
 As previously, we introduce the generating function:
$\displaystyle  K_s(x) = \sum_{n \ge 0} Z_n(x) s^n ,
$
%The expression for the generating function $K_s(x)$ is in this case:
and obtain
\begin{align}
K_s(x) = \frac{Z_0(x)}{1 - s e^{-x} \cosh x} + 4 s e^{x} \sinh x \; h_s(x)^{\frac{1}{-2 + s} } \int_x^{\infty} e^{\frac{2(1-s)y}{2 - s}} h_s(y)^{ -2 + \frac{1}{2 - s}} Z_0(y) dy .
\label{eqKs}
\end{align}
Focusing on the choice $P_0(x) = \delta(x - y)$, we set $Z_0(x) = \left[ \delta(x - y) - \delta(x + y) \right]/2$. For $0 < x < y$, the position of the singularity with the smallest modulus lies at the same position as for the symmetric part: $s = R(x)^{-1}$. It is thus the sub-exponential factors that discriminate the decay rates of the symmetric and anti-symmetric components.
Expanding the generating function to first order around $x=0$ near the pole at $s = 1$, we find:
\begin{align}
  K_s(x) &\simeq  \frac{x (\coth y - 1)}{1 - s} .
  \end{align}
  This gives the expected small $x$ asymptotic behavior for $Z_n(x)$:
  \begin{align}
  Z_n(x) &\simeq x (\coth y - 1) .
\end{align}
This confirms that the antisymmetric component is indeed negligible, in the large $n$ limit, against its symmetric counterpart: 
$|Z_n(x)| \ll S_n(x)$, for small enough $|x|$. 
%component has a $n$ sub-exponential term because of 
This stems from the additional $h_s(x)^{-1}$ factor present in Eq.~(\ref{eqQs}), as compared to Eq.~(\ref{eqKs}).

\end{appendix}

% plain will sort alphabetically, remove it for order of mention in the text
%\bibliographystyle{plain}
\bibliography{mainbib_jphysa.bib}

%merlin.mbs apsrev4-1.bst 2010-07-25 4.21a (PWD, AO, DPC) hacked
%Control: key (0)
%Control: author (8) initials jnrlst
%Control: editor formatted (1) identically to author
%Control: production of article title (-1) disabled
%Control: page (0) single
%Control: year (1) truncated
%Control: production of eprint (0) enabled
\begin{thebibliography}{14}%
\makeatletter
\providecommand \@ifxundefined [1]{%
 \@ifx{#1\undefined}
}%
\providecommand \@ifnum [1]{%
 \ifnum #1\expandafter \@firstoftwo
 \else \expandafter \@secondoftwo
 \fi
}%
\providecommand \@ifx [1]{%
 \ifx #1\expandafter \@firstoftwo
 \else \expandafter \@secondoftwo
 \fi
}%
\providecommand \natexlab [1]{#1}%
\providecommand \enquote  [1]{``#1''}%
\providecommand \bibnamefont  [1]{#1}%
\providecommand \bibfnamefont [1]{#1}%
\providecommand \citenamefont [1]{#1}%
\providecommand \href@noop [0]{\@secondoftwo}%
\providecommand \href [0]{\begingroup \@sanitize@url \@href}%
\providecommand \@href[1]{\@@startlink{#1}\@@href}%
\providecommand \@@href[1]{\endgroup#1\@@endlink}%
\providecommand \@sanitize@url [0]{\catcode `\\12\catcode `\$12\catcode
  `\&12\catcode `\#12\catcode `\^12\catcode `\_12\catcode `\%12\relax}%
\providecommand \@@startlink[1]{}%
\providecommand \@@endlink[0]{}%
\providecommand \url  [0]{\begingroup\@sanitize@url \@url }%
\providecommand \@url [1]{\endgroup\@href {#1}{\urlprefix }}%
\providecommand \urlprefix  [0]{URL }%
\providecommand \Eprint [0]{\href }%
\providecommand \doibase [0]{http://dx.doi.org/}%
\providecommand \selectlanguage [0]{\@gobble}%
\providecommand \bibinfo  [0]{\@secondoftwo}%
\providecommand \bibfield  [0]{\@secondoftwo}%
\providecommand \translation [1]{[#1]}%
\providecommand \BibitemOpen [0]{}%
\providecommand \bibitemStop [0]{}%
\providecommand \bibitemNoStop [0]{.\EOS\space}%
\providecommand \EOS [0]{\spacefactor3000\relax}%
\providecommand \BibitemShut  [1]{\csname bibitem#1\endcsname}%
\let\auto@bib@innerbib\@empty
%</preamble>
\bibitem [{\citenamefont {Cauchy}(1847)}]{Cauchy1847}%
  \BibitemOpen
  \bibfield  {author} {\bibinfo {author} {\bibfnamefont {A.}~\bibnamefont
  {Cauchy}},\ }\href@noop {} {\bibfield  {journal} {\bibinfo  {journal}
  {Applied Mathematical Sciences}\ }\textbf {\bibinfo {volume} {25}},\ \bibinfo
  {pages} {536} (\bibinfo {year} {1847})}\BibitemShut {NoStop}%
\bibitem [{\citenamefont {Robbins}\ and\ \citenamefont
  {Monro}(1951)}]{Robbins1951}%
  \BibitemOpen
  \bibfield  {author} {\bibinfo {author} {\bibfnamefont {H.}~\bibnamefont
  {Robbins}}\ and\ \bibinfo {author} {\bibfnamefont {S.}~\bibnamefont
  {Monro}},\ }\href@noop {} {\bibfield  {journal} {\bibinfo  {journal} {The
  Annals of Mathematical Statistics}\ }\textbf {\bibinfo {volume} {22}},\
  \bibinfo {pages} {400} (\bibinfo {year} {1951})}\BibitemShut {NoStop}%
\bibitem [{\citenamefont {Spall}(2012)}]{Spall2012}%
  \BibitemOpen
  \bibfield  {author} {\bibinfo {author} {\bibfnamefont {J.~C.}\ \bibnamefont
  {Spall}},\ }\enquote {\bibinfo {title} {Stochastic optimization},}\ in\
  \href@noop {} {\emph {\bibinfo {booktitle} {Handbook of Computational
  Statistics: Concepts and Methods}}},\ \bibinfo {editor} {edited by\ \bibinfo
  {editor} {\bibfnamefont {J.~E.}\ \bibnamefont {Gentle}}, \bibinfo {editor}
  {\bibfnamefont {W.~K.}\ \bibnamefont {H{\"a}rdle}}, \ and\ \bibinfo {editor}
  {\bibfnamefont {Y.}~\bibnamefont {Mori}}}\ (\bibinfo  {publisher} {Springer
  Berlin Heidelberg},\ \bibinfo {address} {Berlin, Heidelberg},\ \bibinfo
  {year} {2012})\ p.\ \bibinfo {pages} {173}\BibitemShut {NoStop}%
\bibitem [{\citenamefont {Sra}\ \emph {et~al.}(2011)\citenamefont {Sra},
  \citenamefont {Nowozin},\ and\ \citenamefont {Wright}}]{Sra_ML}%
  \BibitemOpen
  \bibfield  {author} {\bibinfo {author} {\bibfnamefont {S.}~\bibnamefont
  {Sra}}, \bibinfo {author} {\bibfnamefont {S.}~\bibnamefont {Nowozin}}, \ and\
  \bibinfo {author} {\bibfnamefont {S.~J.}\ \bibnamefont {Wright}},\
  }\href@noop {} {\emph {\bibinfo {title} {{Optimization for Machine
  Learning}}}}\ (\bibinfo  {publisher} {The MIT Press},\ \bibinfo {year}
  {2011})\BibitemShut {NoStop}%
\bibitem [{\citenamefont {Frenkel}\ and\ \citenamefont
  {Smith}(2002)}]{FrenkelSmith}%
  \BibitemOpen
  \bibfield  {author} {\bibinfo {author} {\bibfnamefont {D.}~\bibnamefont
  {Frenkel}}\ and\ \bibinfo {author} {\bibfnamefont {B.}~\bibnamefont
  {Smith}},\ }\href@noop {} {\emph {\bibinfo {title} {Understanding Molecular
  Simulations}}},\ \bibinfo {edition} {2nd}\ ed.\ (\bibinfo  {publisher}
  {Adademic Press},\ \bibinfo {year} {2002})\BibitemShut {NoStop}%
\bibitem [{\citenamefont {Newman}\ and\ \citenamefont
  {Barkema}(1999)}]{NewmanBarkema}%
  \BibitemOpen
  \bibfield  {author} {\bibinfo {author} {\bibfnamefont {M.~E.~J.}\
  \bibnamefont {Newman}}\ and\ \bibinfo {author} {\bibfnamefont {G.~T.}\
  \bibnamefont {Barkema}},\ }\href@noop {} {\emph {\bibinfo {title} {Monte
  Carlo Methods in Statistical Physics}}}\ (\bibinfo  {publisher} {Oxford
  University Press},\ \bibinfo {year} {1999})\BibitemShut {NoStop}%
\bibitem [{\citenamefont {Krauth}(2006)}]{Krauth}%
  \BibitemOpen
  \bibfield  {author} {\bibinfo {author} {\bibfnamefont {W.}~\bibnamefont
  {Krauth}},\ }\href@noop {} {\emph {\bibinfo {title} {Statistical Mechanics:
  Algorithms and Computations}}}\ (\bibinfo  {publisher} {Oxford Master Series
  in Physics},\ \bibinfo {year} {2006})\BibitemShut {NoStop}%
\bibitem [{Note1()}]{Note1}%
  \BibitemOpen
  \bibinfo {note} {The $T=0$ limit is not innocuous, for it breaks detailed
  balance \cite {Monthus}}\BibitemShut {NoStop}%
\bibitem [{\citenamefont {Inc.}()}]{Mathematica}%
  \BibitemOpen
  \bibfield  {author} {\bibinfo {author} {\bibfnamefont {W.~R.}\ \bibnamefont
  {Inc.}},\ }\href {https://www.wolfram.com/mathematica} {\enquote {\bibinfo
  {title} {Mathematica, {V}ersion 12.3},}\ }\BibitemShut {NoStop}%
\bibitem [{\citenamefont {Paris}(2013)}]{Paris2013}%
  \BibitemOpen
  \bibfield  {author} {\bibinfo {author} {\bibfnamefont {R.~B.}\ \bibnamefont
  {Paris}},\ }\href@noop {} {\bibfield  {journal} {\bibinfo  {journal} {Applied
  Mathematical Sciences}\ }\textbf {\bibinfo {volume} {7}},\ \bibinfo {pages}
  {6601} (\bibinfo {year} {2013})}\BibitemShut {NoStop}%
\bibitem [{\citenamefont {the authors}(2021)}]{CMST1}%
  \BibitemOpen
  \bibfield  {author} {\bibinfo {author} {\bibnamefont {the authors}},\
  }\href@noop {} {\bibfield  {journal} {\bibinfo  {journal} {in preparation}\ }
  (\bibinfo {year} {2021})}\BibitemShut {NoStop}%
\bibitem [{\citenamefont {Marsaglia}\ \emph {et~al.}(2000)\citenamefont
  {Marsaglia}, \citenamefont {Tsang} \emph {et~al.}}]{marsaglia2000ziggurat}%
  \BibitemOpen
  \bibfield  {author} {\bibinfo {author} {\bibfnamefont {G.}~\bibnamefont
  {Marsaglia}}, \bibinfo {author} {\bibfnamefont {W.~W.}\ \bibnamefont
  {Tsang}},  \emph {et~al.},\ }\href@noop {} {\bibfield  {journal} {\bibinfo
  {journal} {Journal of statistical software}\ }\textbf {\bibinfo {volume}
  {5}},\ \bibinfo {pages} {1} (\bibinfo {year} {2000})}\BibitemShut {NoStop}%
\bibitem [{\citenamefont {Press}\ \emph {et~al.}(2007)\citenamefont {Press},
  \citenamefont {William}, \citenamefont {Teukolsky}, \citenamefont
  {Vetterling}, \citenamefont {Saul},\ and\ \citenamefont
  {Flannery}}]{press2007numerical}%
  \BibitemOpen
  \bibfield  {author} {\bibinfo {author} {\bibfnamefont {W.~H.}\ \bibnamefont
  {Press}}, \bibinfo {author} {\bibfnamefont {H.}~\bibnamefont {William}},
  \bibinfo {author} {\bibfnamefont {S.~A.}\ \bibnamefont {Teukolsky}}, \bibinfo
  {author} {\bibfnamefont {W.~T.}\ \bibnamefont {Vetterling}}, \bibinfo
  {author} {\bibfnamefont {A.}~\bibnamefont {Saul}}, \ and\ \bibinfo {author}
  {\bibfnamefont {B.~P.}\ \bibnamefont {Flannery}},\ }\href@noop {} {\emph
  {\bibinfo {title} {Numerical recipes 3rd edition: The art of scientific
  computing}}}\ (\bibinfo  {publisher} {Cambridge university press},\ \bibinfo
  {year} {2007})\BibitemShut {NoStop}%
\bibitem [{\citenamefont {Flajolet}\ and\ \citenamefont
  {Sedgewick}(2009)}]{acombinatorics}%
  \BibitemOpen
  \bibfield  {author} {\bibinfo {author} {\bibfnamefont {P.}~\bibnamefont
  {Flajolet}}\ and\ \bibinfo {author} {\bibfnamefont {R.}~\bibnamefont
  {Sedgewick}},\ }\href@noop {} {\emph {\bibinfo {title} {Analytic
  Combinatorics}}}\ (\bibinfo  {publisher} {Cambridge University Press},\
  \bibinfo {year} {2009})\BibitemShut {NoStop}%
\end{thebibliography}%

\end{document}